\documentclass[12pt,preprint]{aastex}
\usepackage{mathptmx}
\usepackage{natbib}
\usepackage{epsf}
\usepackage{colordvi}

\shorttitle{Study of current sheets in the wake of two crossing filaments eruption}
\shortauthors{Dai et al.}

\begin{document}

\title{Current sheets in the wake of an eruption of two crossing filaments}

\author{Jun Dai\altaffilmark{1,2,3,4}, Jiayan Yang\altaffilmark{1,2,4*}, Leping Li\altaffilmark{2,3*} and Jun Zhang\altaffilmark{2,3}}

\affil{$^1$Yunnan Observatories, Chinese Academy of Sciences, Kunming 650216, China}
\affil{$^2$University of Chinese Academy of Sciences, Beijing 100049, China}
\affil{$^3$Key Laboratory of Solar Activity, National Astronomical Observatories, Chinese Academy of Sciences,  Beijing 100101, China}
\affil{$^4$Center for Astronomical Mega-Science, Chinese Academy of Sciences, Beijing 100101, China}
\email{yjy@ynao.ac.cn; lepingli@nao.cas.cn}

\begin{abstract}
Employing Solar Dynamic Observatory/Atmosphertic Imaging Assembly (AIA) multi-wavelength images, we study an eruption of two crossing filaments, and firstly report the current sheets (CSs) connecting the lower flare ribbons and the upper erupting filaments. On  July 8, 2014, two crossing filaments are observed in the NOAA active region (AR) 12113. The lower-lying filament rises first, and then meets the higher-lying one. Thereafter, both of them erupt together. The filament eruption draws the overlying magnetic field lines upward, leading to the approach of two legs, with opposite magnetic polarities, of the overlying field lines. Two sets of bright CSs form at the interface of these two legs, and magnetic reconnection takes place in the CSs producing the underneath flare ribbons and post-flare loops.  Several bright plasmoids appear in the CSs, and propagate along the CSs bi-directionally. The CSs and plasmoids are observed in AIA multi-wavelength channels, indicating that both of them have been heated during the reconnection process, with hot and warm components. Employing the differential emission measure (EM) analysis, we find that both the temperature and EM of the CSs decrease from the flare arcades outward to the erupting filaments, and those of the plasmoids are significantly larger than the regions where no plasmoid is detected.

\end{abstract}


\section{Introduction}

Solar filaments (prominences) are chromospheric and coronal structures that contain relatively cool and dense plasma suspended along the magnetic polarity inversion lines (PILs) \citep{Gaizauskas1997A,jiang01}. They are observed as dark ``filaments" on the solar disk, and bright ``prominences" above the solar limb in multi-wavelength images \citep{mack10}. According to their locations, the filaments are usually divided into three types: (1) active region (AR) filaments, (2) intermediate filaments, and (3) quiescent filaments \citep{yan15}. Among them, the AR filaments are usually active and sometimes erupt \citep{chen09,shen12,schm15,yang17}. The successful filament eruptions are closely associated with the coronal mass ejections (CMEs) and flares \citep{zhou03,torok05,chen11,song18}. 

The erupting filament stretches its overlying magnetic field lines upward, makes their legs with opposite polarities converge \citep{shibata99,lin2000JGR}. The current sheets (CSs) then form at the interface, connecting the bottom of the upper erupting filament and the tip of the lower flare arcades \citep{Li2016A,yan18}. Magnetic reconnection takes place in the CSs, converting magnetic energy to other forms, e.g. thermal, kinetic, and particle \citep{reev11,li2016b}.
Following the CME leading edge, the white light (WL) ray-like structures are detected in coronagraph images, and presented as observational evidence of CSs \citep{Webb2003J}. Employing Ultraviolet Coronagraph Spectrometer (UVCS) observations, CSs are identified as narrow hot features of plasma between the flare arcades and CME cores \citep{Ciaravella2002A,Ko2003A,Bemporad2006A,Ciaravella2008A,Schettino2010A}.
Using Solar Ultraviolet Measurements of Emitted Radiation (SUMER) observations, CSs are proposed as hot gas with large line widths above the post-flare loops \citep{Innes2003S,Wang2007A}. Utilizing multi-viewpoint observations, \cite{Patsourakos2011A} analyze a post-CME ray, and suggest that it is indeed related to a post-CME CS. Employing the Solar Dynamics Observatory \citep[SDO,][]{pesn12}/Atmospheric Imaging Assembly \citep[AIA,][]{lemen12} multi-wavelength images, the CSs are detected as very high temperature plasma \citep{reev11,Savage2012A,Li2016A}. Recently, various properties of CSs, including the temperature, width, and density, have also been investigated \citep{Ciaravella2008A,su13,Ciaravella2013A,Sun2014A,Li2016A}.

Crossing filaments, reflecting the complex magnetic field topology nearby, are rarely observed and investigated. How are the magnetic fields distributed near the crossing filaments? How do the surrounding magnetic fields evolve during eruptions of the crossing filaments? These are still open questions. On July 8, 2014, an eruption of two crossing filaments in NOAA AR 12113 was recorded by SDO/AIA. In this Letter, we study the evolution of the filament eruption, and report the CSs connecting the lower flare ribbons and the upper erupting filaments for the first time. The observations and results are separately presented in Sections 2 and 3. The summary and discussion are presented in Section 4.

\section{Data and observations}
SDO/AIA is a set of normal incidence imaging telescopes, acquiring solar atmospheric images in ten wavelength channels. Different AIA channels show plasma at different temperatures. In this study, we employ AIA multi-wavelength ultraviolet (UV) and extreme UV (EUV) images, with spatial sampling and time cadence of 0.6\arcsec\,pixel$^{-1}$ and 12\,s, respectively, to investigate the evolution of the filament eruption and its following CSs.  Helioseismic and Magnetic Imager \citep[HMI,][]{schou12} line-of-sight (LOS) magnetograms, with spatial sampling and time cadence of 0.5\arcsec\,pixel$^{-1}$ and 45\,s, respectively, are also used to analyze the magnetic field evolution of the filament eruption source region. In addition, we employ the WL coronagraph images from LASCO/C2 to study the evolution of the associated CME.

\section{Results}

Two S-shaped filaments, F1 and F2, with lengths of $\sim$37 Mm and $\sim$62 Mm, respectively, are observed in AIA multi-wavelength channels, denoted by green and blue dotted lines in Figure\,\ref{f:eruption}(a). By investigating the structures and evolutions of these two filaments and also the associated photospheric magnetic fields in the source region, we note that they are located upon the complex PILs of the NOAA AR 12113, see Figure\,\ref{f:eruption}(b). Successive flows along the filament F2 are identified, and the filament F2 becomes bright, underneath the filament F1 (see the online animated version of Figure\,\ref{f:eruption}(d)). Two filaments F1 and F2 thus cross each other at almost the middle positions, see Figure\,\ref{f:eruption}(c), and separately connect the southern positive and northern negative polarity magnetic fields of the AR, see Figure\,\ref{f:eruption}(b). Since $\sim$15:50 UT, the lower-lying filament F2 slowly rose, see Figure\,\ref{f:eruption}(c). It met the higher-lying crossing filament F1 at $\sim$16:05 UT. Both of them then erupt together outward to the northeast (see the online animated version of Figure\,\ref{f:eruption}(d)). Along the erupting direction, denoted by a blue line AB in Figure\,\ref{f:eruption}(d), we make a time-slice of AIA 304 \AA~ images, and show it in Figure\,\ref{f:eruption}(e). It indicates that the filaments first rise slowly with a mean speed of $\sim$20 km\,s$^{-1}$, and then erupt quickly with a larger mean speed of $\sim$660 km s$^{-1}$ and a mean acceleration of $\sim$1 km\,s$^{-2}$. Since 16:36 UT, an associated CME, with a mean speed of $\sim$400 km s$^{-1}$, was detected in the FOV of LASCO/C2, see Figures\,\ref{f:eruption}((f)-(g)).

An associated M6.5 class flare took place since $\sim$16:08 UT (see the online animated version of Figure\,\ref{f:flare}). The initial brightenings of two flare ribbons (FRs) are separately enclosed by the blue and red ellipses in Figure 2(a). Thereafter, they elongate bi-directionally along the PILs of the AR, see the green and pink contours in Figure\,\ref{f:flare}(b). Along the elongating directions, denoted by a black line CD in Figure\,\ref{f:flare}(b), we make a time-slice of AIA 1600 \AA~ images, and display it in Figure\,\ref{f:flare}(c). Evident elongations of the FRs are detected at mean speeds of $\sim$40 km s$^{-1}$ northward and 30 km s$^{-1}$ southward, respectively. Furthermore, complex flare ribbons are detected under the crossing parts of two filaments, see Figure\,\ref{f:flare}(b). The post-flare loops appeared since $\sim$16:15 UT, marked by two black dotted lines in Figure\,\ref{f:flare}(d). We overlay the initial brightenings of FRs on Figure\,\ref{f:flare}(d) as blue and red ellipses, and notice that the initial post-flare loops connect the initial FRs. Consistent with the FRs, the brightening of post-flare loops also propagates along the PILs of the AR, see the green and pink contours in Figure\,\ref{f:flare}(e), bi-directionally. Along the propagating direction, denoted by a black line EF in Figure\,\ref{f:flare}(e), we make a time-slice of AIA 171 \AA~ images, and show it in Figure\,\ref{f:flare}(f). The post-flare loops propagate clearly at a mean speed of $\sim$20 km s$^{-1}$. Below the crossing parts of two filaments, similar to the flare ribbons, complex post-flare loops are identified, see Figure\,\ref{f:flare}(e). These observations indicate that the magnetic reconnection in the wake of the erupting filaments occurs in the middle position, and then takes place successively along the AR PILs bi-directionally.

Above the FRs, two sets of bright structures, connecting the lower FRs and the upper erupting filaments, were observed in AIA multi-wavelength channels from 16:12 UT to 16:22 UT (see the online animated version of Figure\,\ref{f:CSs}). We overlay the initial FRs displayed in Figure\,\ref{f:flare}(a) on Figures\,\ref{f:CSs}(a) and (d) as blue and red ellipses. The eastern (western) bright structures root in the eastern (western) initial FRs, rather than the footpoints of the erupting filaments. In the bright structures, several elliptical brighter, i.e., 20\% more intensities than those of the adjacent parts along the bright structure, plasmoids appear and propagate upward and downward along the bright structures. Similar bright ray-like structures are identified in LASCO/C2 coronagraph images in the wake of the erupting filaments, see Figures\,\ref{f:eruption}((f)-(g)). All these observations support that these bright structures represent the post-eruption CSs connecting the lower FRs and the upper erupting filaments.

Figure\,\ref{f:CSs}(a) shows one CS (CS1) in AIA 94 \AA~ channel at 16:13:37 UT with a width, the full width at half maximum of the intensity profile of the AIA 94 \AA~channel centering around, perpendicular to the CS, of $\sim$2.8 Mm. Here, the intensity profile of the AIA 94\,\AA~channel is revised by subtracting the nearby background. By tracing the plasmoids, with similar shape and without sudden change, in the animation frame by frame, we follow the evolution of plasmoids. The triangles in Figures\,\ref{f:CSs}((a)-(c)) mark a plasmoid, P1, in the CS at different times. They indicate that the plasmoid moves upward along the CS. By fitting the temporal evolution of the displacements of plasmoid centroid linearly, we measure the mean moving speed of the plasmoid, and obtain a mean value of 360 km s$^{-1}$, see the green arrow in Figure\,\ref{f:CSs}(b). From 16:14:25 UT, another plasmoid, P2, marked by a square in Figure\,\ref{f:CSs}(c), appeared in the CS1, and moved downward to the solar surface with a mean speed of $\sim$320 km s$^{-1}$, see the green arrow in Figure\,\ref{f:CSs}(d). The CS1 disappeared after $\sim$16:17 UT. The red dashed line in Figure\,\ref{f:CSs}(d) outlines another set of CS (CS2) at 16:14:49 UT, with the width of $\sim$3.5 Mm. Several fine structures, i.e., threads, of CS2 are detected clearly in AIA multi-wavelength channels, e.g. 335, 193, 171, 131, and 304\,\AA, with a mean width of $\sim$1.5\,Mm (see the online animated version of Figure\,\ref{f:eruption}(d)). Moreover, the CS width in coronagraph images is also calculated with a mean value of $\sim$40\,Mm. The circles in Figures\,\ref{f:CSs}((d)-(f)) mark a plasmoid, P3, in the CS2. It moves upward with a speed of $\sim$350 km s$^{-1}$. In this event, most of the plasmoids move outward along the CSs. Due to the filament eruption, the CS2 moves laterally, see the colored dashed lines in Figure\,\ref{f:CSs}(f). The CSs become invisible in AIA wavelength channels after $\sim$16:22 UT, lasting for $\sim$10\,min.

The temperature and EM maps are obtained using the DEM analysis method \citep{Cheng2012A}, where six AIA channels, including 131\,\AA, 94\,\AA, 335\,\AA, 211\,\AA, 193\,\AA, and 171\,\AA, are employed. Figures\,\ref{f:DEM}(a) and (b) separately show the temperature and the EM maps at 16:44 UT, when the CS2 were well-developed and the plasmoids were clearly detected. Along the CS2, the spatial distributions of the mean temperature and EM within the white rectangles in Figures\,\ref{f:DEM}(a) and (b) are investigated, and showed in Figure\,\ref{f:DEM}(c) as black and blue curves, respectively. The temperature drops from $\sim$8 MK to $\sim$5 MK in $\sim$30\arcsec, while the EM decreases from 2.8$\times$10$^{28}$ cm$^{-5}$ to 1.5$\times$10$^{28}$ cm$^{-5}$, along the CS2 outward. Assuming that the LOS depth, \textit{d}, of the CS equals its width (3.5\,Mm), and is constant along the length, we estimate the density, \textit{n}, along the CS from the EM by using \textit{n=(EM\,d$^{-1}$)$^{0.5}$}, and obtain a range of (6.5-8.9)$\times$10$^{9}$ cm$^{-3}$. Several peaks of the temperature and EM curves are detected, corresponding to the plasmoids formed in the CS2. They show larger values than other CS part where no plasmoid is detected. A plasmoid, marked by the blue rectangles in Figures\,\ref{f:DEM}((a)-(b)), is investigated in detail. Its DEM curve is displayed in Figure\,\ref{f:DEM}(d). Here, 100 Monte Carlo realizations of the data are computed. The upper and lower ends of the colored rectangles in Figure\,\ref{f:DEM}(d) can be regarded as estimates of the uncertainties in the best-fit solution, the black solid curve in Figure\,\ref{f:DEM}(d), indicating how well the DEM is determined at a given temperature bin. The fairly broad curve, with mean temperature and EM of 7.8 MK and 9.8$\times$10$^{28}$ cm$^{-5}$, respectively, indicates that the plasmoid shows a multi-temperature structure, consistent with the observation that the CSs and plasmoids are both observed in AIA multi-wavelength channels.

\section{Summary and discussion}

On July 8, 2014, two crossing filaments, located upon the complex PILs of NOAA AR 12113, are observed. The lower-lying filament rises slowly and meets the higher-lying one. Both of them then erupt together, resulting in a CME and an M\,6.5 flare. The FRs and post-flare loops first appear in the middle, and then successively propagate bi-directionally along the AR PILs. Two sets of CSs, connecting the lower initial FRs and the upper erupting filaments, are identified in AIA multi-wavelength images and LASCO/C2 WL coronagraph images. Bright plasmoids appear in the CSs, and move along the CSs upward and downward. They also appear in AIA multi-wavelength channels, showing multi-temperature plasma structures. The temperature and EM of CSs both decrease along the CS outward away from the solar surface.

Based on the AIA multi-wavelength observations, schematic diagrams are provided in Figure\,\ref{f:cartoon} to explain the evolution of the filament eruption and its underlying CSs. Because of the eruption of two crossing filaments, see Figure\,\ref{f:cartoon}(a), the overlying magnetic filed lines are drawn upward. The legs of the overlying field lines, with opposite magnetic polarities, converge and interact. The CSs then form at the interacting region, see Figure\,\ref{f:cartoon}(b). Magnetic reconnection, taking place in the CSs, produces the underlying FRs and post-flare loops, see Figure\,\ref{f:cartoon}(c).

\begin{figure}
\includegraphics[width=0.6 \textwidth]{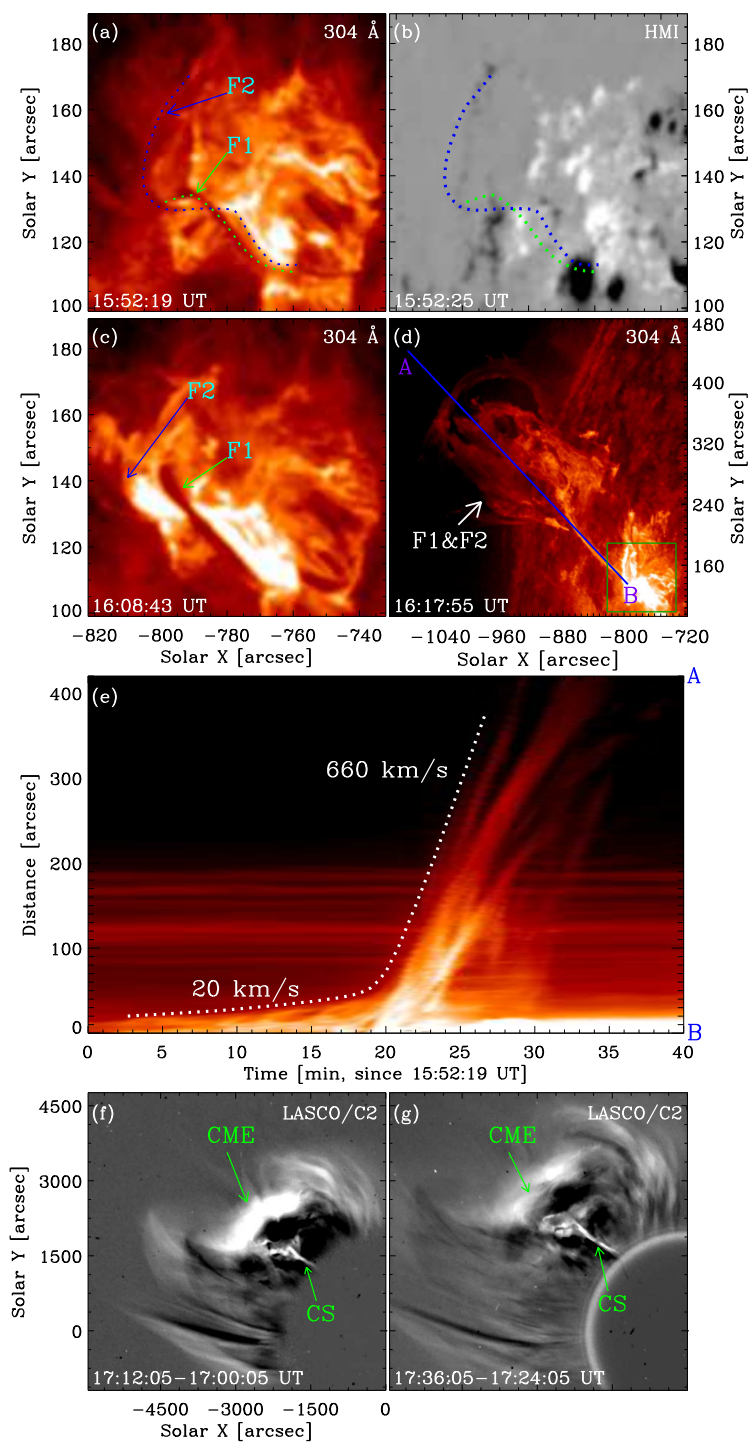}
\caption{\small{Eruption of two crossing filaments. AIA 304 \AA~images ((a), (c), and (d)) and an HMI LOS magnetogram (b). (e) A time-slice of AIA 304 \AA~images along the blue line AB in (d). ((f)-(g)) LASCO/C2 running difference coronagraph images. The green and blue dotted lines in (a) and (b) outline two crossing filaments, F1 and F2, respectively. The green rectangle in (d) shows the field of view (FOV) of ((a)-(c)). The white dotted line in (e) outlines the filament eruption, and the erupting speeds are denoted by the numbers in (e). (An animation of panel (d) is available.)}}
\label{f:eruption}
\end{figure}

\begin{figure}
\includegraphics[width=1.\textwidth]{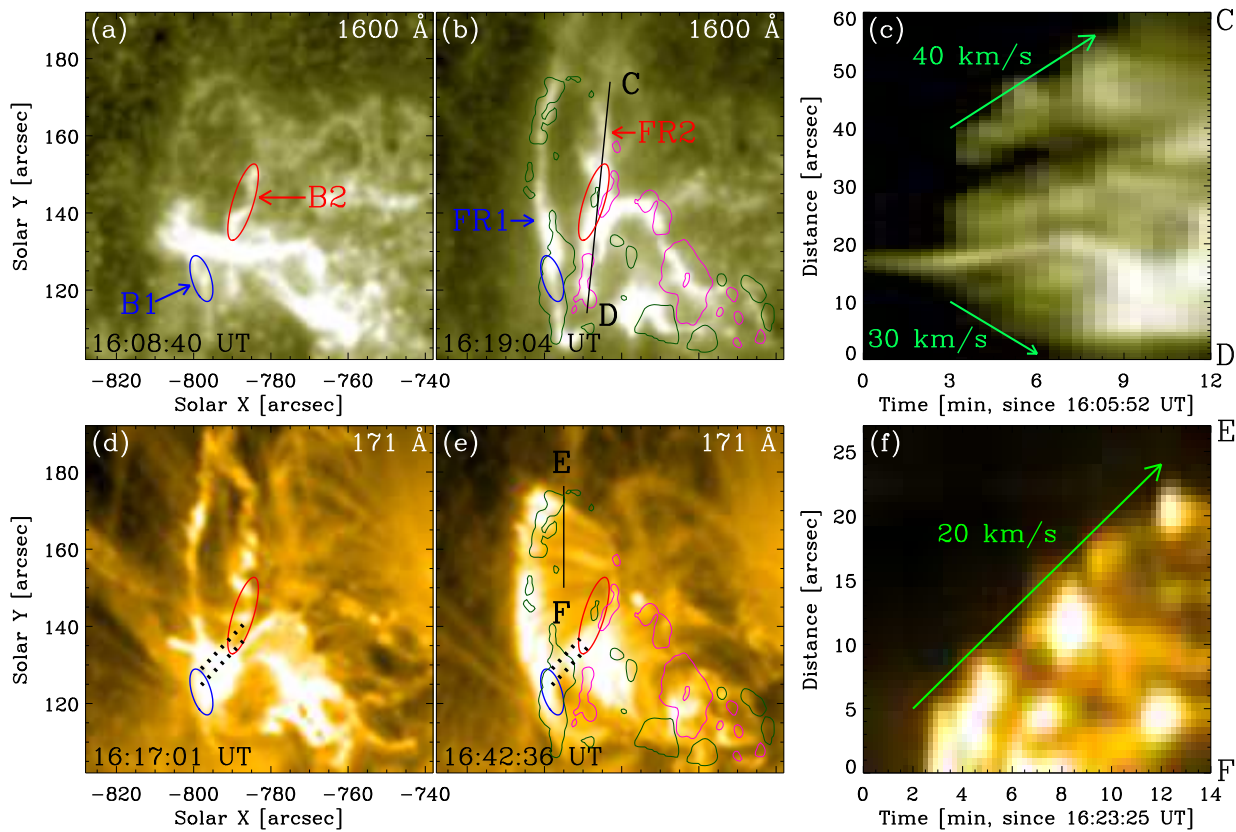}
\caption{\small{Evolution of associated FRs and post-flare loops. AIA 1600\,\AA~((a)-(b)) and 171\,\AA~((d)-(e)) images, and time-slices of AIA 1600\,\AA~(c) and 171\,\AA~(f) images along the lines CD and EF in (b) and (e), respectively. The blue and red ellipses in ((a)-(b) and (d)-(e)) enclose the initial FRs, and the dotted lines in ((d)-(e)) outline the initial post-flare loops. The green and pink contours in (b) and (e) represent the negative and positive polarity magnetic fields surrounding the crossing filaments with magnetic field strengths of -50 and 220 G, respectively. The blue arrows in (c) and (f) represent the movements of FRs and post-flare loops along the AR PILs. The propagating speeds are separately denoted by numbers in (c) and (f). (An animation of this figure is available.)}}
\label{f:flare}
\end{figure}

Crossing filaments are rarely studied, which represent the complex topology of the photospheric magnetic field \citep{Filippov2011S}, see Figure\,\ref{f:eruption}(b). No magnetic reconnection signature, e.g. brightening, is detected between these two crossing filaments when they meet. It indicates that these two filaments have same polarity axial field lines, consistent with the connections between the filaments and the photospheric magnetic fields, see Figure\,\ref{f:eruption}(b). 

The evolution of the FRs and post-flare loops hints essential information of three dimensional (3D) magnetic reconnection during the filament eruption. Both the FRs and post-flare loops appear in the middle, and then propagate along the PILs of the AR. This shows that the magnetic reconnection between two legs of the overlying field lines of the filament eruption takes place first in the middle, and then successively happens along the AR PILs bi-directionally. This evolution of the magnetic reconnection is caused by the nature of the filament eruption, that starts to rise from the middle part \citep{Li2009A}. Furthermore, both the complex flare ribbons and the complex post-flare loops are detected under the crossing parts of two filaments, representing the complex magnetic fields distributed near the crossing filaments.

\begin{figure}
\includegraphics[width=0.8\textwidth]{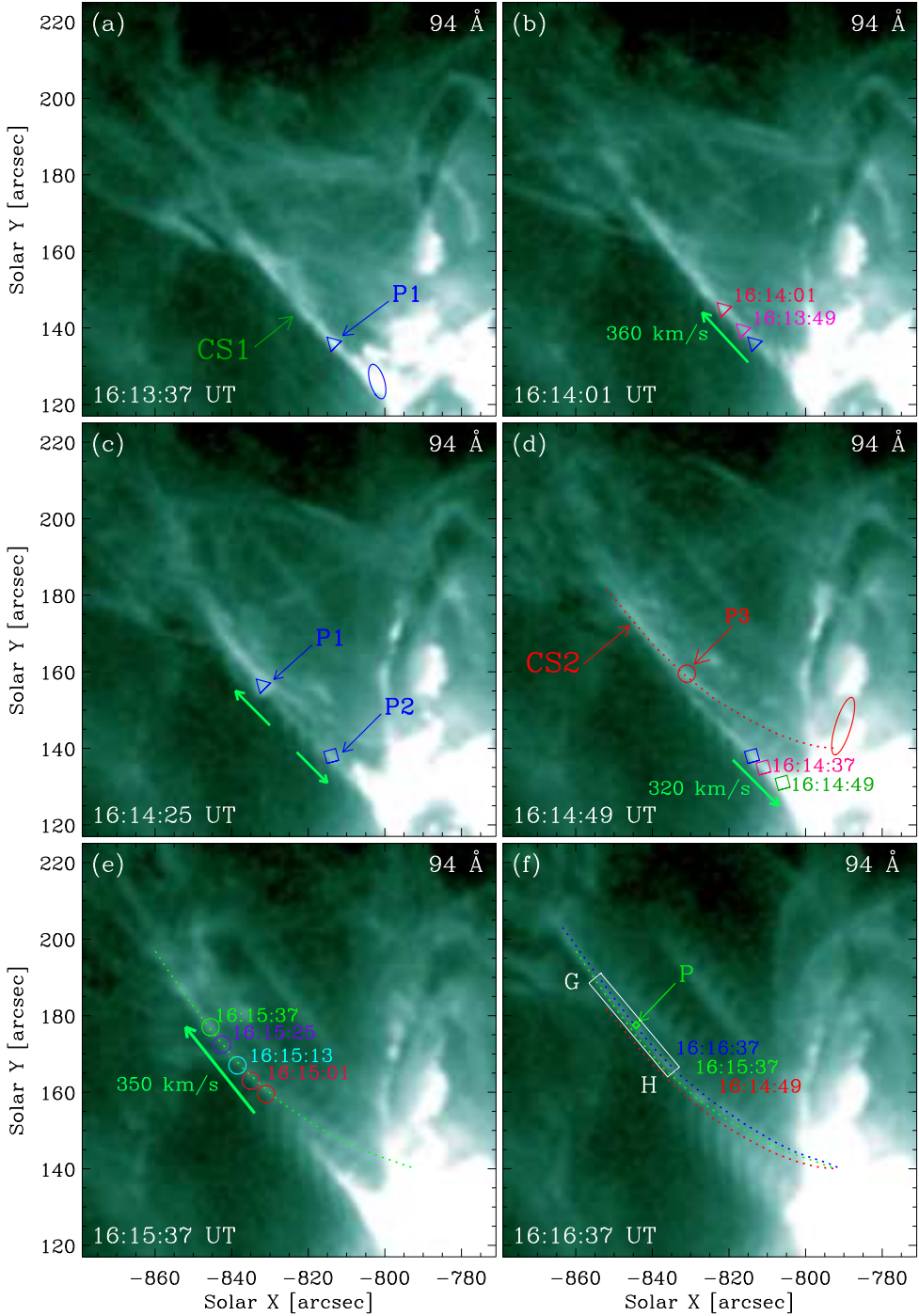}
\caption{\small{Evolution of two sets of CSs. ((a)-(f)) AIA 94 \AA~images. The blue and red ellipses in (a) and (d) enclose the initial FRs as displayed in Figure\,\ref{f:flare}(a). The triangles, rectangles, and circles in ((a)-(e)) mark the plasmoids at different times, and the dotted lines in ((d)-(f)) outline the CSs. The green arrows in ((b)-(e)) denote the moving directions of plasmoids. The moving speeds are denoted by the numbers in the panels. The white rectangle GH in (f) shows the FOV of Figures\,\ref{f:DEM}((a)-(b)). The green rectangle P in (f) marks the plasmoid as showed in Figures\,\ref{f:DEM}((a)-(b)). (An animation of this figure is available.)}}
\label{f:CSs}
\end{figure}

\begin{figure}
\includegraphics[width=0.98\textwidth]{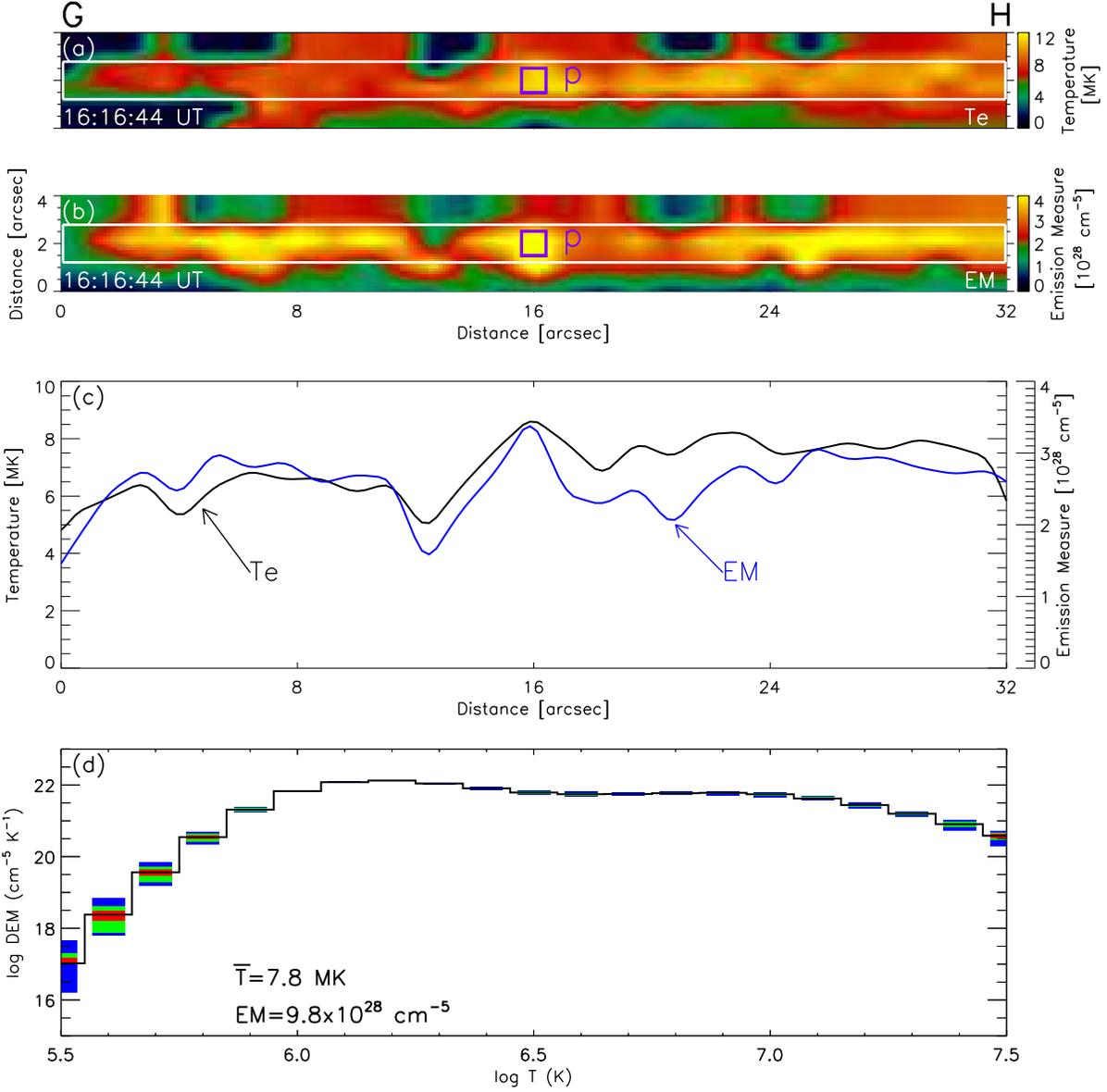}
\caption{\small{Temperature and EM of the CS. ((a)-(b)) Temperature (a) and EM (b) images enclosed by the white rectangle GH in Figure\,\ref{f:CSs}(f). (c) Spatial distributions of the mean temperature (dark curve) and EM (blue curve) of the CS within the white rectangles in (a) and (b). (d) DEM curves of a plasmoid marked by the rectangles in ((a)-(b)). The red, green, and blue rectangles in (d) separately represent the regions containing 50\%, 51-80\%, and 81-95\% Monte Carlo  solutions.}}
\label{f:DEM}
\end{figure}

\begin{figure}
\includegraphics[width=0.98\textwidth]{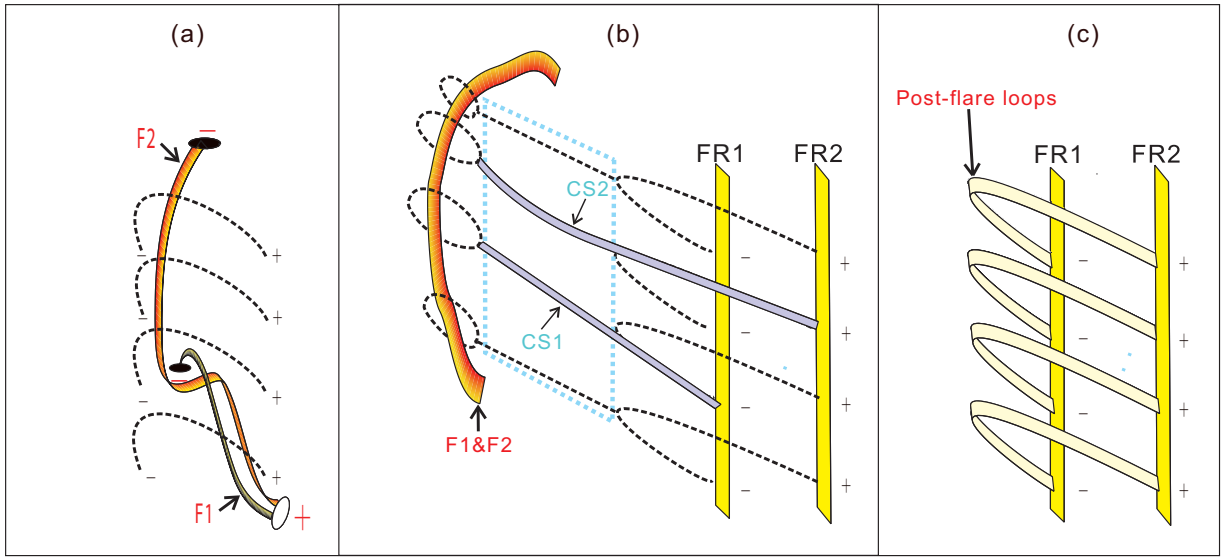}
\caption{\small{Schematic diagrams showing the eruption of two crossing filaments. The pluses and minuses mark the positive and negative polarity magnetic fields. The orange and brown thick lines in (a) and (b) represent two crossing filaments, F1 and F2. The black dotted lines in ((a)-(b)) represent the magnetic field lines overlying the filaments. The blue dotted rectangle in (b) represents the reconnection region, and the gray curves in (b) show the CSs. The yellow lines in (b) and (c) represent the FRs, and the light yellow archs in (c) represent the post-flare loops.}}
\label{f:cartoon}
\end{figure}

The post-eruption CSs are observed on the solar disk, while those post-eruption CSs previously studied are mostly detected above the solar limb \citep{Ciaravella2002A,liu2010A,liuw2013A,Li2016A}. 
For the observations, e.g. WL coronagraph images, away from the solar surface, the filament erupts perpendicular to the LOS in the limb events, and goes out of the FOV of the instruments, e.g. LASCO/C2, quickly. Therefore, only ray-like structures are identified \citep{Ciaravella2002A,Webb2003J,Ciaravella2008A}. However, the filaments in this event erupt radially, and stay a longer time in the FOV of LASCO/C2. Thus the post-eruption CSs connecting the upper erupting filaments, rather than only ray-like structures, are clearly identified, see Figures\,\ref{f:eruption}((f)-(g)). 
For the observations, e.g. EUV images, near the solar surfance, the CSs well above the cusp-shaped flare loops have been reported for the limb events \citep{liu2010A,liuw2013A,Li2016A}. Different from previous studies, the CSs in this study root in the lower FRs, see Figures\,\ref{f:CSs}(a) and (d). This is not reported before. The CSs may include the lower part of the legs of the overlying field lines below the reconnecting region, which constitute the post-flare loops after the magnetic reconnection, see Figure\,\ref{f:cartoon}(b).

The observable structures of CSs are detected depending on the nature, e.g. density and temperature, of plasma in and surrounding the CSs \citep{reev10}. This may be why only two sets of CSs are observed in this event, see Figure\,\ref{f:cartoon}(b). The width of CSs ranges from 1.5-3.5 Mm in EUV images, consistent with \cite{Li2016A}, but smaller than those of \citet{Savage2010A}, \citet{liu2010A}, and \citet{Ling2014A}. It is 40 Mm in WL coronagraph images, identical to those of \citet{Ciaravella2002A}, \citet{Ciaravella2008A}, and \citet{Li2016A}. For different instruments, the CS widths are different may be caused by three reasons: (1) the nature of emissions and wavelength bandpasses; (2) the spatial samplings of different instruments; and (3) the physical mechanisms, e.g. turbance \citep{bemp08,Li2016A}. 

The appearance of plasmoids in the CSs indicates the presence of plasmoid instabilities during the reconnection process \citep{bhat09,Takasao2012d,li2016b}. Both the plasmoids and CSs are observed in AIA multi-wavelength channels, showing that they are heated with multi-temperature plasma. This is consistent with \citet{li2016b}, but different from \citet{hann14} and \citet{Li2016A} as they show only hot CSs. Furthermore, the temperature, EM, and density of the CSs decrease outward from the solar surface along the CSs \citep{Sun2014A,Li2016A}. 

\acknowledgments

We thank the anonymous referee for many valuable comments that helped us to improve the paper substantially. The authors are indebted to the SDO team for providing the data. The work is supported by the National Foundations of China (11633008, 11703084, 11673034, 11533008, 11790304, and 11773039), and Key Programs of the Chinese Academy of Sciences (QYZDJ-SSW-SLH050).

\end{document}